\documentclass[sn-mathphys]{sn-jnl}

\usepackage{amsmath}
\usepackage{cleveref}
\usepackage{amsfonts}
\usepackage{algpseudocode}
\usepackage{mathtools}

\newtheorem*{proposition}{Proposition}

\theoremstyle{definition}

\begin{document}

\title{Modeling routing problems in QUBO with application to ride-hailing}

\author*[1,2]{\fnm{Michele} \sur{Cattelan}}

\author[1]{\fnm{Sheir} \sur{Yarkoni}}

\affil[1]{\orgdiv{Volkswagen Data:Lab}, \orgname{Volkswagen AG}, \orgaddress{\street{Ungererstra\ss e 69}, \city{Munich}, \postcode{80805}, \country{Germany}}}

\affil[2]{\orgdiv{Institute for Theoretical Physics}, \orgname{University of Innsbruck}, \orgaddress{\city{Innsbruck}, \postcode{A-6020}, \country{Austria}}}

    

\abstract{
Many emerging commercial services are based on the sharing or pooling of resources for common use with the aim of reducing costs. Businesses such as delivery-, mobility-, or transport-as-a-service have become standard in many parts of the world, fulfilling on-demand requests for customers in live settings. However, it is known that many of these problems are NP-hard, and therefore both modeling and solving them accurately is a challenge. Here we focus on one such routing problem, the Ride Pooling Problem (RPP), where multiple customers can request on-demand pickups and drop-offs from shared vehicles within a fleet. The combinatorial optimization task is to optimally pool customer requests using the limited set of vehicles, akin to a small-scale flexible bus route. In this work, we propose a quadratic unconstrained binary optimization (QUBO) program and introduce efficient formulation methods for the RPP to be solved using metaheuristics, and specifically emerging quantum optimization algorithms. }

\maketitle

\section{Introduction}
Solving combinatorial optimization problems by mapping them to quadratic unconstrained binary optimization (QUBO) problems is a well-studied field in literature~\cite{Kochenberger2014, lucas2014ising}. The topic has gained increased attention recently in light of advances in quantum optimization, and consequentially, other quantum- and physics-inspired metaheuristic optimization algorithms~\cite{farhi2000quantum, farhi2014quantum, kadowaki1998quantum}. Furthermore, rapid development in the field of quantum technologies has produced prototype quantum processors for public use which are able to execute these quantum algorithms, making them interesting tools to test the power of quantum computing. Companies such as Amazon, Google, D-Wave Systems, IBM, among others, all have cloud-accessible processors that have been examined by researchers and practitioners alike from all around the world~\cite{johnson2011quantum, arute2019quantum, li2021co}. The promise of quantum computing lies in the theoretical demonstration of some quantum algorithms being able to outperform their classical counterparts for particular classes of problems~\cite{arute2019quantum, shor1999polynomial}. Instead of classical bits in binary programming which can only be $0$ or $1$ deterministically, quantum computers (QCs) are built from \textit{quantum bits}, or qubits, which have probabilities of being measured as $0$ and $1$. These qubits allow for the construction of new classes of algorithms where quantum mechanics are exploited to perform tasks that are fundamentally inefficient for classical computers to perform~\cite{feynman2018simulating}. Specifically for optimization, quantum computing researchers have started examining well-known hard optimization problems in hopes of finding quantum algorithms that can utilize the potential of QCs to gain performance advantages relative to classical optimization algorithms. Currently, the most promising such algorithms are quantum annealing (QA)~\cite{kadowaki1998quantum} and the quantum approximate optimization algorithm (QAOA)~\cite{farhi2014quantum}, two metaheuristic quantum optimization algorithms motivated by the adiabatic theorem~\cite{farhi2000quantum}. Both of these algorithms have been implemented in quantum hardware and tested in academic and industrial settings over a variety of optimization problems~\cite{yarkoni2021quantum,bpsp,paintshop_multicar,harrigan2021quantum}. Solving optimization problems using these algorithms requires problems to be formulated either as QUBOs or Ising models, which are isomorphic and NP-hard to minimize~\cite{barahona1982computational, Karp1972}. Furthermore, many canonical NP-hard and NP-complete problems have straightforward transformations to QUBO or Ising~\cite{lucas2014ising}, making them of practical use for application development\footnote{The field of quantum computing hardware and quantum optimization algorithms is a rapidly-evolving field, the technical aspects of which are beyond the scope of this paper. For more information about quantum hardware and quantum optimization, we refer the reader to~\cite{li2021co}.}. 
The limiting factor in formulating optimization problems using state of the art quantum hardware is the number of qubits and the degree of connectivity between them. Therefore, optimization problems with high connectivity between variables requires more layers in quantum circuits (meaning, deeper circuits), or an increase in the number of physical qubits used to represent problem variables, both of which are scarce resources. So, finding minimal QUBO representations that most efficiently use the limited resources can be of significant practical use. \\

In this paper we focus on a particular class of NP-hard combinatorial optimization problems, routing problems, which are of practical relevance to modeling many real-world problems, from logistics and transportation, to drilling holes in circuit boards, to planning of factory floors. While the theoretical motivations for the Traveling Salesperson Problem (TSP), and by extension the Vehicle Routing Problem (VRP), are well-known, modeling complex routing problems often requires application-specific constraints to ensure that solutions are practically feasible. For example, constraints such as service time-windows, vehicle capacities, or real-time traffic conditions all need to be modeled correctly such that the minimum of the QUBO corresponds to the global optima of the original problem. In our work, we investigate one such real-world routing problem, namely the Ride Pooling Problem (RPP), which is formally described as follows: given a fleet of vehicles $A$ and a set of requests $(s,f)$ with \textit{pickup points} $s$ and \textit{drop-off points} $f$ for $n$ customers, assign routes to vehicles in $A$ to pick up and drop off all customers such that the cost of the target (objective) function is minimized. Historically, the RPP had been used to solve flexible fleet-services problems, such as fuel delivery scheduling and optimization~\cite{ke2020ride}. However, the investigation of RPP has gained renewed interest in recent years through the emergence of car-sharing services such as Uber and Lyft, and is therefore of practical interest to both model and solve accurately in practice. As such, objective functions incorporating minimizing lateness of arrivals, waiting time, and similar constraints, are used when solving the RPP. In this work, we focus on solving the RPP using distance minimization of vehicles in the fleet to allow a straightforward motivation (and comparison) with TSP and VRP QUBO models. The basic methods developed in this paper can extend to almost any of these objective functions, and therefore are general enough to be used for a wide variety of applications. Specifically, we show how to construct QUBOs that accurately represent the RPP problem through the inclusion of problem-specific constraints, compare this to canonical representations of TSP and VRP variants in QUBO, and estimate the resources required to solve such RPP QUBOs using quantum processors and quantum algorithms. We demonstrate how the methods we develop could lead to more resource efficient problem representations for quantum optimiziation algorithms. \\

The paper is organized as follows. In \cref{sec:prev}, we review previous papers about QUBO formulations of routing problems primarily motivated by quantum algorithms. Following this, in \cref{sec:QUBO} we present an in-depth analysis of QUBO formulations for routing problems, and show that some choices of binary/decision variables yield impractical QUBO models for metaheuristic optimization algorithms. Finally, in \cref{sec:RPP}
we motivate the constraints required to accurately model the RPP, and present our QUBO formulation of the problem.


\section{Previous works}\label{sec:prev}
We briefly present previous works where TSPs/VRPs (and similar problems) were modeled as QUBOs specifically for quantum optimization (or similarly inspired) algorithms. In~\cite{lucas2014ising}, straightforward representations of Hamiltonian paths, cycles, and therefore TSPs were introduced for binary optimization with either quantum or similar such metaheuristic optimization algorithms. The set of binary decision variables used for all these encode whether each location was visited at a particular step in the path, thus requiring $O(N^2)$ variables for an $n$-location TSP problem (or $n$ node graph, for Hamiltonian path/cycle problems in general). Feld et al.~\cite{Feld2019CVRP} extended this work to the Capacitated VRP (CVRP), showing how to construct a VRP QUBO with capacities. The TSP can be easily extended to VRP using $O(kN^2)$ (where $k$ is the number of vehicles) QUBO variables using the same decision variables as in~\cite{lucas2014ising}. To include the capacity constraints, the authors transform inequalities to equalities with slack variables for QUBO using terms from the knapsack QUBO formulation in~\cite{lucas2014ising}. The authors explore the trade-offs between modeling parts of the problem with QUBOs independently versus solving the CVRP as a single combined QUBO. Similarly, additional work from~\cite{Borowski2020VRPTW} in this direction further showed how to construct CVRP QUBOs from known benchmark data. Another constraint, the \textit{time-window}, was developed for QUBO routing problems in~\cite{Papalitsas2019TSPTW}. The authors use these time-windows such that specific locations in the TSP problem need to be visited within certain times so that a solution is feasible.

\section{Routing problems in QUBO}\label{sec:QUBO}
Quadratic unconstrained binary optimization (QUBO) is defined as follows: given a quadratic cost matrix $Q$ ($N\times N$ real-values), find the minimum $0-1$ assignment to binary vector $x$ (of length $N$) that minimizes the QUBO cost function. Formally, we say that the objective of the QUBO is:
\begin{equation}
    \min_{x} x^T Q x.
\end{equation}
The QUBO problem presented here is NP-hard to minimize in the worst case~\cite{Karp1972}, and furthermore, QUBO is isomorphic to the NP-hard Ising model~\cite{lucas2014ising} under a change of variable. Ising models are a well-studied model from physics, and use spin vectors $s = \{-1, 1\}^N$ to formulate the objective function (represented by a so-called Hamiltonian $H$):
\begin{equation}
    H(s) = \sum_{i} h_is_i + \sum_{i<j} J_{ij}s_is_j.
\end{equation}
Here, $s_i \in s$ are the spin variables, $h_i$ are the associated linear cost terms and $J_{ij}$ are the quadratic interaction terms. Thus, the task is to find a set of spin values such that the cost function (in physics known as the energy of the system) is minimized. Note that in the QUBO case, the linear terms are the diagonals of the matrix $Q$, as $x_i^2 = x_i,  \forall x_i \in \{0, 1\}$. In literature, the Ising model and QUBO are often used interchangeably depending on the problem being formulated. For example, the NP-hard MaxCut problem is usually defined by the Ising model~\cite{barahona1988application}:
\begin{equation}
    \mathrm{MaxCut} = \min \sum_{i<j}J_{ij}s_is_j.
\end{equation}
However, other problems are formulated as QUBOs due to the simplicity of representing certain constraints. For example, the one-hot constraint over a set of binary variables $x_i \in x$ in QUBO is formulated as:
\begin{equation}
    \sum_{i} x_i = 1 \iff \min \left(1-\sum_{i}x_i\right)^2=0.
\end{equation}
The right-hand side of the implication is then added to the objective function (with a high penalty) to ensure that the global minimum of the QUBO is reached when exactly one binary variable $x_i=1$. Graph problems, routing problems, and similar optimization problems all require such constraints, and therefore are usually formulated as QUBOs rather than Ising models. For the remainder of our work, we operate strictly on binary variables to build our optimization problems using QUBOs, although the techniques and results presented hold equally for Ising models using the binary-to-spin variable change of basis.  

As mentioned, routing problems are of particular interest for industrial optimization problems due to their wide applicability and have been studied extensively in literature. To motivate our work, we start by reviewing known QUBO formulations of routing problems. The Traveling Salesperson Problem (TSP) is a well-studied NP-hard combinatorial optimization problem with known approximation results~\cite{christofides1976worst}. Formally, TSP is defined as follows: given a weighted graph $G = (V,E)$ with $n =\mathrm{card}(V)$ nodes (representing locations), find a cycle such that every node is visited exactly once with the minimum sum of weights of edges used (distances and connections between locations). This is also known as a (minimum weight) Hamiltonian cycle. To create the QUBO for TSP, we present the formulation described in \cite{lucas2014ising}. The binary variables we use represent whether the location $v\in V$ is visited as the $i$-th location in the cycle:
\begin{equation*}
    x_{v,i}=\left\{\begin{array}{cc}
         1 & \text{if location $v$ is the $i$-th location to visit},  \\
         0 &\text{otherwise.} 
    \end{array}\right.
\end{equation*}
Note that, in order to represent a cycle, we assume all sums over indices are taken to be modulo $n$. We now use these binary variables to define the objective function and constraints of the TSP QUBO. The objective function of the TSP is the sum of weights along the edges used in the graph; this is typically referred to as the distance. Letting $w_{v,k}$ be the weight of the edge $(v,k)$ in $G$, the distance minimization can be represented as:
\begin{equation}\label{eq:TSPof}
    H_{A}=\sum_{(v,k)\in E}\sum_{i=0}^{n-1}\frac{w_{v,k}}{W} x_{v,i}\,x_{k,i+1},
\end{equation}

where $W:=\epsilon + \max_{(v,k)\in E}w_{v,k}$ (where $\epsilon > 0$) is a normalization factor to bound the weights strictly between 0 and 1. We now continue to formulate the problem as a QUBO by enumerating the constraints. As explained above, in order to define a QUBO objective to minimize, the implementation of the constraints has to be included in the objective function directly. Notice that $H_{A}$ alone is not sufficient, as we can easily see by inspection that the optimal solution is the zero vector with minimum value $0$, since we are not constraining the values that $x_{v,i}$ can take. Therefore, we add the following constraints based on our binary variable definition: only one location can be visited at each step in the tour, and no location is visited twice during a tour. These two conditions result in two separate sets of one-hot constraints, which can be re-written for QUBO as follows: 


\begin{equation}\label{eq:TSPpos_const}
    \hspace{-.225cm} H_{B} = \sum_{v\in V}\left(1-\sum_{i=1}^{n}x_{v,i} \right)^{2} + \sum_{i=1}^{n}\left(1-\sum_{v\in V}x_{v,i} \right)^{2}.
\end{equation}
The two summations in \cref{eq:TSPpos_const} are minimized when the two described constraints are valid. Thus, the global minima of the function given by the sum of \cref{eq:TSPof} and \cref{eq:TSPpos_const} are the tours that minimize both the distance function and satisfy the constraints defined above. Lastly, we implement the constraint that a tour of the locations is a valid Hamiltonian cycle of the graph. This happens only if the cycle is a subgraph of $G(V,E)$, meaning, we must forbid steps in the tour between nodes not connected in $G$. 
In order to include this in the QUBO, we add a penalty term between all decision variables representing non-adjacent nodes in $G$:
\begin{equation}\label{eq:TSPnonedge}
    H_{C}=\sum_{(v,w)\notin E}\sum_{i=1}^{n}x_{v,i}\,x_{w,i+1}.
\end{equation}
From the normalization in \cref{eq:TSPof}, quadratic terms using QUBO variables representing adjacent nodes in $G$ have lower cost than non-adjacent nodes, and therefore the condition of being a Hamiltonian cycle of the original graph is fulfilled. By summing all the contributions in \cref{eq:TSPof}, \cref{eq:TSPpos_const}, and \cref{eq:TSPnonedge} we obtain the QUBO formulation for TSP:
\begin{equation}\label{eq:TSPQUBO}
    H_{TSP}=H_{A}+H_{B}+H_{C}.
\end{equation}


A well-known generalization of the TSP is the Vehicle Routing Problem (VRP). This problem, which is also NP-hard and widely studied in literature, describes the case where a fleet of vehicles (starting from a common \textit{depot}) must each construct a cycle such that a set of locations is only visited once and each vehicle returns to the depot at the end, subject to some minimization function. Therefore, TSP is a special case of VRP where the number of vehicles is one. 

We can formulate VRP as a QUBO by reusing many of the parts introduced for TSP-- we do this by considering a separate TSP sub-problem for each vehicle and apply global constraints to ensure the validity of the solution. Again, consider the graph $G = (V,E)$, where nodes are the set of all the locations (plus the depot), and weighted edges are distances between locations. For generality, we consider the case where $G$ is complete, meaning every location can be reached from any other location. Thus, the binary decision variables are:
\begin{equation*}
    x_{a,v,s} = \left\{\begin{array}{cc}
        1 & \text{the vehicle $a$ is in location $v$ at step $s$,} \\
        0 & \text{otherwise.}
    \end{array} \right.
\end{equation*}
The index $a$ enumerates the vehicles in the fleet, where $A$ is the total number of vehicles. The indices $s$ and $i$ play the same role as in \cref{eq:TSPQUBO}, where again we consider the sum over location indices modulo $n$ (without loss of generality we define the first location $s=0$ to be the depot).


Now, the objective function for the VRP QUBO is:
\begin{equation}\label{eq:VRPof}
    H_{A}^{\text{VRP}}=\sum_{a=1}^{A}\sum_{(v,k)\in E}\sum_{s=0}^{n}\frac{w_{v,k}}{W}x_{a,v,s}\,x_{a,k,s+1},
\end{equation}
with $W$ as defined for TSP. As we can easily see, if we consider the case where $A=1$, we recover the objective function described in \cref{eq:TSPof} for TSP. For the rest of the constraints, i.e. the generalizations of \cref{eq:TSPpos_const}, we have:
\begin{equation}\label{eq:VRPcons}
    H_{B}^{\text{VRP}}=\sum_{v\in V\setminus\{d \}}\left(1-\sum_{a=1}^{A}\sum_{s=0}^{n} x_{a,v,s} \right)^{2}
    +\sum_{a=1}^{A}\sum_{s=0}^{n}\left(1-\sum_{v\in V} x_{a,v,s} \right)^{2}.
\end{equation}

The one-hot constraints are the same as before, but with the addition (in the first summation) that the constraint sums over all vehicles; meaning, that every location can only be visited exactly once, shared between all vehicles. The only location excluded from this constraint is the depot-- since we don't know \textit{a priori} how long the optimal tours of each vehicle are (only that it is upper bounded by the number of locations), we must allow them to ``stay'' in the depot as long as they need to. The second part of \cref{eq:TSPpos_const} ensures that for each vehicle (the outer sum), only one location is visited at a time, as for TSP. For the final QUBO, the VRP QUBO is obtained by summing together \cref{eq:VRPof} and \cref{eq:VRPcons}:
\begin{equation}\label{eq:VRPham}
    H^{\text{VRP}}=H_{A}^{\text{VRP}}+ H_{B}^{\text{VRP}},
\end{equation}
Again note that if we restrict the number of vehicles in the problem to one, then we recover the full TSP QUBO, resulting in correct QUBO representations for both VRP and TSP. 

\subsection{Alternative formulations of routing problems in QUBO}

In the previous section we briefly reviewed the representation of VRP and TSP in QUBO. However, it is important to note that encoding valid constraints in QUBO is inherently dependent on the choice of decision variables. In previous literature, two such possibilities have been described, notably the \textit{node-based} decision variables and the \textit{edge-based} ones. In the node-based version for TSP, decision variable $x_{ij}$ denotes location $i$ is visited at step $j$, with an additional index denoted the vehicle number for VRP. This is the convention we adopted in the previous sections of this paper. For node-based QUBOs, the number of variables is $O(kN^2)$ for $k$ vehicles and $n$ locations. In the edge-based formulations, the same notation is used to describe a related but fundamentally different choice: $x_{ij}$ denotes that the arc between location $i$ and $j$ is used in a tour. Then, constraints are formulated using these variables to ensure each arc is used exactly once. Thus, the number of variables required for the edge-based QUBO is $O\left(\mathrm{card}(E)\right) = O(N^2)$. Note that, in principle, upon using the edge-based formulation, it is necessary to encode constraints such that independent subtours (closed loops) cannot be present in the optimum of the problem. For TSP/VRP heuristics which use tour augmentation, arc insertion, or similar steps in generating solutions, this is not a bottleneck~\cite{dantzig1954solution}; any additional arc which creates a closed loop and violates the constraints is rejected, stored, and eliminated from the candidate list of possible solutions. However, metaheuristics have no such knowledge of the problem constraints, and so all constraints need to be modeled mathematically and incorporated into the objective function directly for the QUBO. Therefore, each possible subtour must be explicitly excluded from the minimum of the objective function, otherwise the optimum becomes dominated by infeasible solutions. Moreover, subtour elimination is $O(2^n)$, and is therefore at least as hard as solving TSP/VRP to begin with. Therefore, to avoid subtour elimination entirely, we have to introduce an ordering of all the edges traversed by each vehicle and formulate constraints such that only valid tours are admissible. This can be done as for the node-based formulation, where each vehicle has a separate set of edge-based decision variables, and each are constrained globally ($k$ arcs are used to leave the depot, $k$ arcs to return, each location is connected to exactly two arcs, etc.). This results in a well-defined QUBO with at most $O(kN^3)$ variables. Obviously, this is asymptotically more variables than the node-based representation of the same problem, and so we conclude that edge-based formulations of routing problems are inappropriate for QUBO\footnote{Previous works which only use explicit subtour elimination in the QUBO for metaheuristic optimization are not referenced in this paper.} solving with metaheuristics.

\section{The Ride-Pooling Problem QUBO}\label{sec:RPP}
The ride-pooling problem is another variant of routing problems that, although less studied than TSP/VRP, is extremely relevant for real-world problems~\cite{molenbruch2017typology,ho2018survey,ke2020ride}. We restate the definition of the RPP for clarity: given a set of ride requests and a fleet of vehicles, pick up all customers and deliver them to their respective drop-off locations while minimizing the distance traveled by the fleet. While qualitatively similar to TSP/VRP, the RPP is unique in two ways: the vehicles do not start at a depot (and are not required to return to a depot), and multiple customers can be picked up consecutively. In practice, additional constraints such as time windows, minimizing customer waiting time, minimizing vehicle deviations, and other such considerations are all included when solving RPPs. 
To construct the QUBO for the RPP, we follow the methods as for VRP and TSP and start by defining the binary decision variables. As before, the variables represent the possible locations at each step of the path for each vehicle in the fleet:
\begin{equation*}
    x_{a,l,\beta} = \left\{\begin{array}{cc}
        1 & \text{vehicle $a$ is in location $l$ at step $\beta$,} \\
        0 & \text{otherwise.}
    \end{array} \right.
\end{equation*}
Index $a$ enumerates the vehicles (up to $A$), $l$ is the index of possible locations, and $\beta$ is the total number of steps in the path for each vehicle. Since every customer has a pickup and drop-off point, the, we have a maximum of $2C+1$ steps in each vehicle's path (including the starting point), where $C$ is the number of ride requests. 

To describe the constraints of the RPP, we start with reusing as much of the formulation from VRP/TSP as possible. Let each ride request be denoted by the tuple $(s_i, f_i)$, where $s_i$ is the $i$th customer's start location (pickup) and $f_i$ is the corresponding final location (drop-off). Therefore, each $(s_i, f_i)$ pair corresponds to two possible locations for each vehicle in the fleet, each of which must be visited by only one vehicle. We denote this set of shared locations for vehicles by $\mathcal{L}$ (note that this does not include the starting point for any of the vehicles). We express the one-hot constraint for each request over all vehicles as follows:   
\begin{equation}\label{cons:location} 
    \sum_{l\in\mathcal{L}}\left(  1- \sum_{a=1}^{A}\sum_{\beta=1}^{S}x_{a,l,\beta} \right)^2 = 0.
\end{equation}
Note that, since we did not include the vehicles' starting points (which we will refer to as $d_a$) in this one-hot constraint, it is of no cost for a vehicle to ``remain'' in that location from one step of its path to the next. This is the same as in the VRP QUBO where vehicles can remain in their depots for as long as necessary, and for the same reason-- we can't know \textit{a priori} how many steps the optimal tour of each vehicle will be in the RPP. So, given that $A$ is the number of vehicles and $S=2C+1$ is the maximum path length, we have that every location can only be visited once in the minimum of this function, excluding the vehicles' starting point. 

Similarly, we constrain all steps in the path such that only one location be visited at a time, also as in TSP/VRP. However, note that, in the case where we have more than one vehicle, the fact that we are guaranteed that at least one location is visited by any one of the vehicles means that at least one location does not need to be visited by any other vehicle. This potentially reduces the lengths of the tours of all vehicles by one step, except in the event that one vehicle must respond to all requests. Thus, we can convert the one-hot constraint corresponding to the last step in the path to a half-hot constraint, where the minimum is when exactly one or no location is visited. This has the additional consequence of making the starting location $d_a$ of each vehicle redundant in the last step, and so we are able to reduce the number of variables in the QUBO without sacrificing feasibility, i.e. we can set $x_{a,d_{a},S}=0$ for all $a$\footnote{Because every request is split into two stops, all path lengths must be even for all the vehicles in the fleet in any feasible solution. This may imply that there is a further improvement to be made (meaning, perhaps we can relax the last two stops of the tour instead of just one), but how to implement this was not immediately obvious, and so it is beyond the scope of the presented QUBO.}. The resulting constraints are therefore a split of the summations in \cref{eq:VRPcons}, a combination of one-hot and half-hot constraints: 

\begin{equation}\label{cons:position}
    \sum_{a=1}^{A}\sum_{\beta=1}^{S-1}\left( 1- \sum_{l\in V} x_{a,l,\beta} \right)^2
    +\sum_{a=1}^{A}\left(1 - 2\sum_{l\in \mathcal{L}} x_{a,l,S}\right)^2 = 0.
\end{equation}
Now, recall that in order for a request to be satisfied, the same vehicle visiting a pickup must also visit the corresponding drop-off. Formally, we say that a request $(s_i,f_i)\in \mathcal{C}$, where $\mathcal{C}$ is set of customer requests, is satisfied if and only if there exist a solution with vehicle $\bar{a}\in\{1,\ldots,A\}$ and steps $0<\beta_{1} < \beta_{2}<S$ such that $x_{\bar{a},s_i,\beta_{1}}=1$ and $x_{\bar{a},f_i,\beta_{2}}=1$. This means that the locations $s$ and $f$ must appear in the same vehicle's tour, and that location $s_i$ must appear before location $f_i$ in that tour, which we call the \textit{causality condition}. We therefore introduce the notion of \emph{incentive terms}: these are penalty terms in the QUBO with negative coefficients for the purpose of incentivizing combinations of binary variables in optimal solutions. To implement the causality condition in the RPP QUBO, we must incentivize all solutions where variable $s_i$ appears before $f_i$ for each vehicle $a$:
\begin{equation}\label{cons:flow&time}
    \sum_{a=1}^{A}\;\sum_{(s,f)\in\mathcal{C}}\;\sum_{\substack{\beta_{1},\beta_{2}=1,\\ \beta_{1}<\beta_{2}}}^{S} -x_{a,s,\beta_{1}}\, x_{a,f,\beta_{2}}.
\end{equation}
Notice that, instead of using such incentive terms, the same result could have been achieved by implementing penalty terms on all possible infeasible configurations of $s_i$ and $f_i$ that violate our causality condition instead. We observe that, for each pair $(s, f) \in \mathcal{C}$, incentivizing feasible or penalizing infeasible solutions yield the same desired result. This is due to the fact that for each feasible solution we know that the value of \cref{cons:flow&time} is $-C$ by definition. Similarly, the set of penalty terms required to penalize all infeasible solutions is the complement to~\cref{cons:flow&time}:
\begin{equation*}
    \sum_{a=1}^{A}\;\sum_{(s,f)\in\mathcal{C}}\;\sum_{\substack{\beta_{1},\beta_{2}=1,\\ \beta_{1}\ge\beta_{2}}}^{S} x_{a,s,\beta_{1}}\, x_{a,f,\beta_{2}} + \sum_{a_1\neq a_2}\;\sum_{(s,f)\in\mathcal{C}}\;\sum_{\beta_{1},\beta_{2}=1}^{S} x_{a_1,s,\beta_{1}}\, x_{a_2,f,\beta_{2}},
\end{equation*}
which by definition evaluates to 0 for feasible solutions. Since the sets containing the terms of the two functions are complementary, and for each $(s, f) \in \mathcal{C}$ the union of the two is all combinations of binary variables in which $s$ and $f$ appear (together with the one-hot constraints), we therefore only require the smaller set to include in the RPP QUBO. We formalize this observation with the following proposition. 
\begin{proposition}
The causality condition can be fulfilled by either incentivizing all feasible solutions or penalizing all infeasible solutions, and whichever uses fewer terms in the QUBO is sufficient.
\end{proposition}

We note that the RPP shares many characteristics with other well-known NP-hard optimization problems formulated as QUBOs. This suggests that our proposition can be applied in general in constrained combinatorial optimization problems modeled as QUBOs and is not unique to RPP. Specifically, our causality condition exploits the property of distinguishing between feasible and infeasible solutions to a constrained optimization problem by penalization, which is more general than the routing application we apply it to. However, generalizing to a theorem requires more general definitions and assumptions over the model and therefore is out of scope for this paper. \\


For the objective function of the RPP, similar to TSP/VRP, we minimize the weight of the arcs between locations. In order to accommodate the modifications in \cref{cons:position} where $d_a$ is removed from the last step, we modify the objective function to match:
\begin{equation}\label{eq:RPPof}
    \sum_{a=1}^{A}\sum_{l_{1}\in V}\sum_{l_{2}\in\mathcal{L}}\sum_{\beta=1}^{S-1}\frac{w_{l_{1},l_{2}}}{W}\,x_{a,l_{1},\beta}\,x_{a,l_{2},\beta+1} +\sum_{a=1}^{A}\sum_{l\in\mathcal{L}}\sum_{\beta=1}^{S-2}\frac{w_{d,l}}{W}\,x_{a,l,\beta}\,x_{a,d_{a},\beta+1}.
\end{equation}
Here, the first addend sums the distances between every location (\textit{including} $d_a$) to all other locations (\textit{excluding} $d_a$) in subsequent steps in the each vehicle's path. The second addend adds the distance terms between all locations to $d_a$ in subsequent steps, except for the last step of the tour. Note that the cost of a vehicle ``staying'' in $d_a$ from one step to the other is 0. Again, by our definition of normalization factor $W$, it is never favorable to either violate constraints nor to return to $d_a$ after leaving it. Thus, by summing \cref{cons:location}, \cref{cons:position}, \cref{cons:flow&time}, and \cref{eq:RPPof}, we obtain the basic RPP QUBO formulation. 
\subsection{Adding real-world constraints to the RPP QUBO}
To make our QUBO more representative of real-world problems, we can add additional complexity in the form of constraints or requirements which must be fulfilled in order to consider solutions feasible (or even optimal, in the case of multi-objective optimization). For example, consider that some real-world ride-hailing companies have a fleet with vehicles of different passenger capacities, or that requests can be made by groups with different numbers of people to be picked up. To solve this problem we must include the concept of capacity for vehicles in the RPP QUBO. Let each vehicle have a fixed capacity $\Omega_{a}$, and each request have a number of passengers $p_{l}$ that need to be collected. Then, at every step of the path, for every vehicle, the number of passengers in a vehicle cannot exceed the vehicle's capacity. This can be implemented via the inequality:
\begin{equation*}
    \sum_{l\in\mathcal{L}}\sum_{\beta=1}^{S}p_{l}x_{a,l,\beta}\le \Omega_{a}\quad\text{for}\; a=1,\ldots,A.
\end{equation*}
Note that $p_{s_i} = -p_{f_i}$. Meaning, the number of passengers entering the vehicle at every pickup $s_i$ is the same as the number of passengers exiting the vehicle at $f_i$\footnote{Note that, in the case where pickups are in separate locations but drop-offs are the same across multiple requests (or vice versa), we can always separate these into two consecutive steps in the path with no distance between them so that this condition always holds.}. To implement these inequalities in QUBO, we require a set of slack variables (denoted by $y_{a,c}$) for each vehicle. This in effect keeps a running sum at each step of the path for every vehicles ensuring that the vehicle capacity is never violated. We express the constraint as the following equality:
\begin{equation}\label{cons:capacity}
    \sum_{a=1}^{A}\left(\sum_{\beta=1}^{S}\sum_{j=0}^{\beta}\sum_{l\in\mathcal{L}}p_{l}x_{a,l,j}-\sum_{c=1}^{\Omega_{a}}y_{a,c} \right)^{2} = 0.
\end{equation}

Thus, we have a complete description of the RPP including vehicle capacities. Note that many similar such real-world constraints can be implemented in similar ways, as has been done in previous literature for TSP/VRP QUBOs. We briefly present a complexity analysis of the RPP QUBO model by enumerating the number of variables needed in order to implement it. Since each vehicle is treated the same, we count the variables on each ``slice'' independently and then multiply by the number of vehicles $A$. The maximum number of steps in a vehicle's path is $2C+1$. Therefore, the number of variables for each vehicle is $(2\cdot C+1)^{2}-1 $, since at each step the vehicle can visit any location except $d_a$ in the last step. Thus, the total number of variables scales as $O(A\cdot C^{2})$ for the basic RPP. 

Although the number of variables required for the basic RPP QUBO scales quadratically, and therefore polynomially, for quantum computing this still represents a significant overhead due to the limited number of qubits that are available in state-of-the-art processors at the time of writing. Indeed, for each variable in the QUBO model we need at least one qubit in the quantum computer. Hence, finding ways to lower the number of variables in the QUBO can have an impact on quantum algorithm performance. By relying on some intuition from the description of the RPP, we can remove some redundant variables by fixing their values \textit{a priori}. For example, each vehicle has a unique starting point, which implies that no other location can be the first step in the path. This means that, for all locations $\mathcal{L}$ we can fix the variables with $\beta=1$ as follows:
\begin{equation*}
    x_{a,d_{a},1}=1\;\text{and}\; x_{a,l,1}=0\quad\text{for all}\; a=1,\ldots, A\;\text{and}\; l\in\mathcal{L}.
\end{equation*}
Furthermore, by analyzing constraints in \cref{cons:flow&time}, we see that it is impossible to go from $d_a$ directly to a drop-off. Furthermore, it is impossible to end a path at any pickup. We can fix these variables as well:
\begin{equation*}
    x_{a,s,S}=0\;\text{and}\; x_{a,f,2}=0\quad\text{for all}\; a=1,\ldots,A\;\text{and}\; (s,f)\in\mathcal{C}.
\end{equation*}

Although this is a modest (linear) improvement in the number of variables used, adopting this in practice would allow practitioners to solve larger sets of RPP QUBOs than without this method. However, to make these QUBOs more representative of real-world problems, we must also include additional constraints in the problem, and account for them in the number of variables in the QUBO. Here we use the capacities of the vehicles as an example. The number of QUBO variables increases by the number of slack variables needed to represent the capacity constraints with equalities. By definition of slack variables, they must sum to the capacity of the vehicle, and hence we require $\Omega_a$ variables for each step in the path. Thus, we need at most $O(A\mathcal{P}C)$ additional QUBO variables for all capacity constraints, where $\mathcal{P}\coloneqq\max_{a=1,\ldots, A}\Omega_{a}$. In total, the number of variables for the RPP QUBO with capacity constraints scales as $O(A[C^{2}+\mathcal{P}])$. 


\section{Acknowledgements}
MC and SY are funded by the German Ministry
for Education and Research (BMB+F) in the project
QAI2-Q-KIS under grant 13N15587. Furthermore, the authors would like to thank Gabriele Compostella, Anestis Papanikolaou, Andrea Skolik and Anton Suchaneck for valuable discussions and suggestions given.

\bibliography{lib}

\appendix

\section{Edge-based QUBO formulation of RPP}
As mentioned in \cref{sec:QUBO}, edge-based formulations of routing problems increase of the number of variables with respect to node-based formulations, and do not seem to be suitable for QUBOs and metaheuristics. However, for consistency, we present an edge-based formulation of the RPP that yields some interesting insights. We start by defining the binary variables in the problem:
\begin{equation*}
    x_{a,i,j,\beta}=\left\{\begin{array}{cc}
        1 & \text{if vehicle $a$ goes from $i$ to $j$ at step $\beta$,} \\
        0 & \text{otherwise,} 
    \end{array} \right.
\end{equation*}
where $a=1,\ldots, A$, $\beta=1,\ldots, S-1$, $i\in V$ and $j\in\mathcal{L}$. The first difference we notice is that the objective function can be easily implemented as following:
\begin{equation}\label{edge:of}
    \sum_{a=1}^{A}\sum_{\beta=1}^{S-1}\sum_{i\in V}\sum_{j\in\mathcal{L}} \frac{w_{ij}}{W}\, x_{a,i,j,\beta},
\end{equation}
where $w_{ij}$ and $W$ are defined as in \cref{sec:RPP}. Although the objective function is composed of only linear terms, we see that we require more variables than in the node-based QUBO. However, one benefit is that, from the definition of the constraints of the RPP, we can construct the edge-based QUBO by using only sums of binomials instead of the incentive terms in \cref{cons:flow&time}. We modify \cref{cons:location} and \cref{cons:position} according to our new variable definitions, meaning we must ensure that each pickup and drop-off location is only entered and left once. Let $\mathrm{source}[l]$ and $\mathrm{target}[l]$ be the arcs entering and exiting location $l$, respectively. Then, for all pickups we require:
\begin{multline}\label{edge:pick}
    \sum_{s:(s,f)\in\mathcal{C}}\left[\left(1-\sum_{a=1}^{A}\sum_{\beta=1}^{S-1}\; \sum_{i\in\text{source[s]}}x_{a,i,s,\beta} \right)^{2}\right. +\\ +\left. \left(1-\sum_{a=1}^{A}\sum_{\beta=1}^{S-1}\; \sum_{j\in\text{target[s]}}x_{a,s,j,\beta} \right)^{2}\right] ,
\end{multline}
and for all drop-offs:
\begin{equation}\label{edge:drop}
    \sum_{f:(s,f)\in\mathcal{C}}\left(1-\sum_{a=1}^{A}\sum_{\beta=1}^{S-1}\; \sum_{i\in\text{source[f]}}x_{a,i,f,\beta} \right)^{2}.
\end{equation}
Now we impose that the starting point must be left by each vehicle, as follows:
\begin{equation}\label{edge:depot}
    \sum_{a=1}^{A}\left(1-\sum_{\beta=1}^{S-1}\;\sum_{j\in\text{target}[d_{a}]} x_{a,d_{a},j,\beta} \right)^{2}.
\end{equation}
Lastly, we impose that a pickup must be visited before its drop-off, and both by the same vehicle. Using the edge-based representation, we can implement the constraint as a sum of binomial terms:
\begin{multline}\label{edge:time&flow}
    \sum_{a=1}^{A}\,\sum_{(s,f)\in\mathcal{C}}\;\sum_{\substack{\beta_{1},\beta_{2},\beta_{3}=1\\\text{s.t.}\,\beta_{2}>\beta_{1},\\\beta_{3}\ge\beta_{2}}}^{S-1}\left[\sum_{i\in\text{source}[s]}\;\sum_{j\in\text{target}[s]}(x_{a,i,s,\beta_{1}}+x_{a,s,j,\beta_{2}})\right.+\\
    -2\left.\sum_{i\in\text{source}[f]}x_{a,i,f,\beta_{3}} \right]^{2}.
\end{multline}

We mentioned in \cref{sec:QUBO} that edge-based formulations of a routing problem can contain some infeasible subtours. We show now that this is not the case by our explicit enumeration of steps in the path. Let us consider a solution $x$ of the edge-based RPP. We show that every path for each vehicle starts in its depot $d_a$\footnote{Note that by fixing the variables with $\beta=1$, as we did in \cref{sec:RPP}, we have an easy reformulation of the problem without these variables. Since it is a trivial generalization, we skip it.}. We cannot start in a pickup point $s$ because otherwise the first addend in sum \cref{edge:pick} is not minimized, and hence a constraint is violated. Furthermore, we cannot start in a drop-off $f$ as well otherwise \cref{edge:drop} can never be satisfied for that $f$. Therefore the starting point must be $d_a$. If we now look at the variables with $\beta=2$, we see that it must be a pickup. Otherwise, \cref{edge:time&flow} is violated because the variables $x_{a,i,f,\beta_{3}}$, with $i\ne d_{a}$ and $\beta_{3}>2$, are all zero, and therefore the request $(s,f)\in\mathcal{C}$ contributes with at least one non-zero term. Now, by considering any pickup $s$ we see that it cannot be visited twice, since if $s$ is entered twice then \cref{edge:pick} is violated. Furthermore, a variable $x_{a,i,j,S-1}$ cannot have $j=s$, otherwise the first sum of \cref{edge:pick} is violated, and, since $s$ cannot appear twice in the same position, it means that the only way for a pickup to be visited in a feasible path is if it is not an end point of the path. Lastly, a drop-off cannot appear twice, otherwise \cref{edge:drop} is violated. Therefore our edge-based RPP QUBO results in feasible solutions that are simple paths that start in each $d_a$, pass through some pickups and drop-offs, and end in a drop-off point.

We immediately see that this formulation of the basic RPP QUBO is not favorable in comparison to the node-based QUBO in \cref{sec:RPP}. The number of variables here scale cubically in the number of ride requests $C$, which is asymptotically worse.
\end{document}